\documentclass[12pt]{article}

\IfFileExists{srcltx.sty}{\usepackage[active]{srcltx}}

\usepackage{amssymb,amsmath,bm,array}%
\usepackage{graphicx}%

\usepackage{xspace,paralist} %
\usepackage[a4paper,hscale=.72,vscale=.72,nohead]{geometry} %

\usepackage[colorlinks=true,bookmarks=false]{hyperref}

\IfFileExists{hypernat.sty}{\usepackage{hypernat}}

\let\oldbox=\Box %
\renewcommand{\Box}{\mathop{\oldbox}}

\let\oldint=\int %
\renewcommand{\int}{\mathop{\hskip -.25ex\oldint \hskip -.4ex}}

\newcommand{\0}{{(0)}} %
\newcommand{\cs}{\textsc{cs}} 

\newcommand{\p}[1]{\partial_{#1}} %
\newcommand{\pz}{\partial_{z}} %
\newcommand{\la}{\langle} %
\newcommand{\ra}{\rangle} %
\newcommand{\parfrac}[2]{\left(\frac{#1}{#2}\right)} %
\newcommand{\Dsl}{\mathop{\lefteqn{D}{\,\mbox{\large /}}}\nolimits} 

\newcommand{\g}[1]{\ensuremath{\gamma^{#1}}}

\renewcommand{\u}[1]{\ensuremath{\mathrm{U}(#1)}\xspace} 
\newcommand{\vb}{\ensuremath{v_{B}}} %
\newcommand{\f}[2]{\ensuremath{F_{#1#2}}} %
\newcommand{\h}[1]{\ensuremath{\bm{H_{#1}}}} %
\newcommand{\ax}{\ensuremath{A_x}\xspace} 
\renewcommand{\b}[1]{\ensuremath{B_{#1}}\xspace} %
\newcommand{\bz}{\ensuremath{B_{z}}\xspace} %
\newcommand{\fb}{\ensuremath{F^{B}}\xspace} %
\newcommand{\fboz}{\ensuremath{F^{B}_{0z}}\xspace} %
\newcommand{\fbl}{\ensuremath{F_{B}}} %

\renewcommand{\k}{\bm{\kappa}} %
\newcommand{\mb}{\ensuremath{m_{B}}}%
\newcommand{\mf}{\ensuremath{m_\Psi}}%
\newcommand{\qf}{\ensuremath{e_{\Psi}}}%
\newcommand{\ma}{\ensuremath{m_{a}}}%
\newcommand{\mg}{\ensuremath{m_{\gamma}}\xspace}%
\newcommand{\mh}{\ensuremath{m_{\gamma\,H}}}

\newcommand{\ev}{~\text{eV}} %
\newcommand{\kev}{~\text{keV}} %
\newcommand{\gev}{~\text{GeV}} %
\newcommand{\lab}{\text{lab}} %
\renewcommand{\star}{\text{star}} %
\newcommand{\CL}{\mathcal{L}} %

\newcommand{\ffd}{(F\tilde F)}  

\begin{document}

\title{\hfill \hbox{\small CERN-PH-TH/2007-135}\\[\baselineskip]
  Anomaly induced effects in a magnetic field}%
\author{\large \sf Ignatios~Antoniadis~$^{a,}$\thanks{On leave from CPHT (UMR
    CNRS 7644) Ecole Polytechnique, F-91128 Palaiseau}~, ~~Alexey
  Boyarsky~$^{a,}$\thanks{On leave of absence from Bogolyubov Institute of
    Theoretical Physics, Kyiv, Ukraine}~,
  ~~Oleg Ruchayskiy~$^{b}$\\[0.5cm]
  $^a$\normalsize\emph{Department of Physics, CERN - Theory Division, 1211
    Geneva 23,
    Switzerland}\\
  $^b$\normalsize\emph{Ecole Polytechnique F\'ed\'erale de
    Lausanne, Institute of Theoretical Physics}\\
  \normalsize\emph{FSB/ITP/LPPC, BSP 720,
    CH-1015, Lausanne, Switzerland}} %
\date{}%
\maketitle
\begin{abstract}
  We consider a modification of electrodynamics by an additional light massive
  vector field, interacting with the photon via Chern-Simons-like coupling.
  This theory predicts observable effects for the experiments studying the
  propagation of light in an external magnetic field, very similar to those,
  predicted by theories of axion and axion-like particles.  We discuss a
  possible microscopic origin of this theory from a theory with non-trivial
  gauge anomaly cancellation between massive and light particles (including,
  for example, millicharged fermions). Due to the conservation of the gauge
  current, the production of the new vector field is suppressed at high
  energies.  As a result, this theory can avoid both stellar bounds (which
  exist for axions) and  the bounds from CMB considered recently, allowing for
  positive results in experiments like ALPS, LIPPS, OSQAR, PVLAS-2, BMV, Q\&A,
  etc.
\end{abstract}

\section{Introduction}
\label{sec:introduction}

A number of experiments, studying the properties of vacuum in a strong
external magnetic field (in particular the propagation of light in such a
field), have been performed over the recent years, or are being in operation
at the moment
\cite{BFRT:92,BFRT:93,PVLAS:05a,PVLAS:05b,QandA:06,PVLAS:07a,OSQAR:06,%
  ALPS:07,BMV:06,Robilliard:07,Afanasev:06}.  One of the main motivations for
these experiments is the search for a hypothetical particle, the axion, which
was originally predicted~\cite{Weinberg:77,Wilczek:77} by the Peccei-Quinn
mechanism of solving the ``strong CP problem'' in QCD
\cite{Peccei:77a,Peccei:77b}. The main property of this particle from the
laboratory experiments' point of view is the coupling of the axion field
$a(x)$ to the photon via the interaction term $\frac14a(x)F\tilde F=a(x)(\vec
E \cdot \vec H)$.  More generally, axion-like-particles (ALP) are
pseudo-scalars which interact with $F\tilde F$,\footnote{We denote by $\tilde
  F$ the dual field strength: $\tilde F_{\mu\nu} = \frac12
  \epsilon_{\mu\nu\lambda\rho} F^{\lambda\rho}$.}
but may not have other
properties of the QCD axion needed to solve the strong CP problem.

In~{\cite{Antoniadis:06} and~\cite{anomaly-th,anomaly-exp} another class of
  effective theories, which also predict effects in the presence of electric and
  magnetic fields, having completely different particle physics motivation,
  were considered.
The coupling of axion to $F\tilde F$ appears in general from chiral
\emph{gauge anomalies}, that may be present in theories with chiral couplings
of fermions to a gauge field. Although gauge anomaly makes theory
inconsistent, the theories with several chiral fields may be well-defined if
the overall anomaly cancels and the overall gauge current is conserved.  The
fact that the net vacuum current is conserved does not mean that it is zero.
If it is not, the non-trivial anomaly cancellation between heavy and light
fields may give rise to experimental signatures of the heavy fields which, due
to the topological nature of anomalies, are not suppressed by the mass of the
heavy fermions and, therefore, may be observed at low
energies~\cite{anomaly-exp}. For theories with non-trivial electromagnetic
anomaly cancellations, such effects will be proportional to $F\tilde F$.
    
    In this paper, we discuss in more details the theory considered
    in~\cite{Antoniadis:06} and, in particular, show how the creation of the
    new light particles, present there, may be suppressed with energy and,
    therefore, the parameters of the theory may not be constrained by the
    stellar bounds, which severely restrict the parameter space of axions~(for
    a recent review, see e.g.~\cite{Raffelt:06a}). We describe the possible
    origin of the effective Lagrangian of~\cite{Antoniadis:06} and show that
    it may appear, for example, from a theory of millicharged fermions,
    interacting with the photon and a ``paraphoton'' similar to the theory
    considered in~\cite{Gies:06a,Ahlers:06}. This allows to have the effects
    of vacuum dichroism and birefringence~\cite{Maiani:86,Raffelt:88} and
    ``shining light through the wall''~\cite{VanBibber:87}, observable in
    experiments like PVLAS~\cite{Roncadelli:07}, OSQAR~\cite{OSQAR:06},
    ALPS~\cite{Ringwald:06,ALPS:07}, BMV~\cite{BMV:06,Robilliard:07},
    LIPPS~\cite{Afanasev:06} avoiding not only the stellar constraints for the
    Chern-Simons like interaction of the paraphoton, but also the bounds on
    the parameters of millicharged particles discussed
    in~\cite{Melchiorri:07}.
    
    The organization of the paper is as follows. In
    Section~\ref{sec:effective-theory}, we review the effective field theory
    of~\cite{Antoniadis:06} that introduces a new massive vector boson,
    coupled to the photon via a Chern-Simons term, and study the propagation
    of light in a magnetic field. At low energies, this theory behaves like a
    theory of ALP corresponding to the longitudinal component of the new gauge
    field. In Section~\ref{sec:avoid-stell-constr}, we modify it at high
    energies, so that the gauge boson couples to a conserved current,
    suppressing the production of its longitudinal component in stars. This
    modification is parametrized by effective non-local terms.  We then
    provide an example of a microscopic theory reproducing these terms, using
    millicharged fermions. We also analyze the various experimental constrains
    and deduce the allowed parameter space that leaves open the possibility of
    detecting interesting effects in forthcoming experiments, such as
    measuring dichroism and birefringence of light and photon regeneration in
    the presence of strong magnetic fields. Section~\ref{sec:discussion}
    contains our concluding remarks and a discussion on possible derivation of
    our model from D-brane constructions.  

\section{Effective theory at low energies and propagation of light in the
  magnetic field.}
\label{sec:effective-theory}

We start by reminding the model considered
in~\cite{Antoniadis:06}. We consider the effective theory of two vector
fields: the usual photon $A_\mu$ and a massive field \b\mu (paraphoton):\footnote{To simplify the notations and avoid proliferation of
  non-essential $\epsilon$-tensors, we will often use differential forms, such as
  $A\wedge D\theta \wedge F_A$ instead of $\frac 12
  \epsilon^{\mu\nu\lambda\rho} A_\mu (\p\nu\theta+\b\nu) F_{\lambda\rho}$ and
  $F\wedge F$ instead of $\frac 14 \epsilon^{\mu\nu\lambda\rho} F_{\mu\nu}
  F_{\lambda\rho}$.}
\begin{equation}
  \label{eq:4}
    S_{low} = \int d^4x \,\left( -\frac14 F_A^2 -\frac14 F_B^2 +  \frac{\mb^2}2 (D_\mu\theta)^2 + \frac{\mg^2}2 (D_\mu\chi)^2 + 2\k D\chi \wedge
  D\theta \wedge F_A    \right)
\end{equation}
where $D\theta = d\theta + B$, $D\chi = d\chi + A$ and $\k$ is a dimensionless
coupling. The Stuckelberg field
$\theta$ makes the action explicitly gauge invariant under $U(1)_B$ gauge
transformations, provided that $\theta\to\theta-\alpha$ when $\b\mu \to \b\mu
+ \p\mu\alpha$.  The field \b\mu is therefore massive with mass \mb.
Another Stuckelberg field $\chi$, transforming as $\chi \to \chi-\lambda$ when
$A_\mu \to A_\mu + \p\mu\lambda$, restores the gauge invariance with respect
to the $U(1)_A$ gauge group.  This means that the photon $A_\mu$ is also
massive with mass \mg. This becomes explicit in the \emph{unitary gauge}
($\theta=0$, $\chi = 0$):
\begin{equation}
   \label{eq:9}
    S_{low,\, unitary} = \int d^4x\, \left(-\frac14 F_A^2 -\frac14 F_B^2 +
    \frac{\mb^2}2 
    \b\mu^2 + \frac{\mg^2}2 A_\mu^2 + 2\k A \wedge
  B \wedge F_A    \right)
\end{equation}

Notice, that sending $\mg\to 0$ (and making the field $\chi$ non-dynamical)
will make $A_\mu$ massless and add the constraint $F_A\wedge F_B = 0$ (with
the field $\chi$ being a Lagrange multiplier). To make the analysis simpler,
we will assume massless photon and the action~(\ref{eq:9}) supplemented by
this constraint, although our analysis remains unchanged in the case of finite
but small enough \mg.  The Chern-Simons (CS) term $\k A\wedge D\theta \wedge
F_A$ remains gauge invariant in this case.
Notice, that in the theory~(\ref{eq:4}) the field \b\mu couples to a
non-conserved current $(J_B^\mu)_{low} $:
\begin{equation}
  \label{eq:7}
  (J_B^\mu)_{low} = \frac{\delta S_{low}}{\delta\b\mu} = \k \epsilon^{\mu\nu\lambda\rho} A_\nu
  F_{\lambda\rho}\quad;\quad \p\mu (J_B^\mu)_{low} = \frac\k2\epsilon^{\mu\nu\lambda\rho} F_{\mu\nu}
F_{\lambda\rho}
\end{equation}
The conserved (N\"other) current is a linear combination of $(J_B^\mu)_{low}$
and $\p\mu\theta$.  As we will show below, the property~(\ref{eq:7}) implies
that in the theory~(\ref{eq:4}) the longitudinal part of the massive vector
field \b\mu behaves as a massive axion with the mass \mb\ and the coupling
constant $\mb/\k$.

\subsection{Propagation of light in a magnetic field}
\label{sec:prop-light-magn}

Let us consider the equations of motion of theory~(\ref{eq:4}) for the case of
the propagation of photons in а strong external magnetic field.  We consider
light propagation along the $z$-axis in the magnetic field $\h0 =
F_{yz}=F_{23}$, pointing in $x$-direction. Under the condition $F_A\wedge F_B
= 0$, the CS term does not depend on the choice of the background vector
potential. For simplicity we choose the background vector potential,
corresponding to the field \h0, in the form $\bar A_y = -z\h0$. Working in the
unitary gauge (i.e. $\theta=0$) for the $B$-field, we obtain the
following system of equations:
\begin{align}
\label{eq:220}
\p\nu F^{\mu\nu} & = -\k\epsilon^{\mu\nu\lambda\rho} \Bigl( 2 B_\nu
F_{\lambda\rho} -
A_\nu \fb_{\lambda\rho}\Bigr)\\
\partial_\nu \fbl^{\mu\nu} - \mb^2 B^\mu &=
\k\epsilon^{\mu\nu\lambda\rho}A_\nu F_{\lambda\rho}
\label{eq:248}\\
\epsilon^{\mu\nu\lambda\rho}F_{\mu\nu}\fb_{\lambda\rho} &= 0\label{eq:250}
\end{align}
One can construct the solution of these equations as perturbations in small
parameter $\k$ around the ``zeroth order approximation'': $A^\0_\mu = a_\mu
e^{i\omega(t-z)}$, $\b\mu^\0 = 0$.

A simple analysis shows that only the longitudinal part of \b\mu (i.e.
longitudinal electric field $\fboz$) interacts with the propagating photon,
with polarization parallel to the external magnetic field.  To demonstrate the
similarities and differences between this model and the axion, we revert from
eqs.~(\ref{eq:220}-\ref{eq:248}) to the system, involving the longitudinal
degree of freedom of the $B$-field and the photon \ax.\footnote{We choose the
  gauge in which the plane wave, with polarization vector parallel
  to the external magnetic field is described by the potential $\ax(t,z)$.} %
Namely, one can represent the field \b\mu in terms of one scalar degree of freedom, $\phi$:
\begin{equation}
  \label{eq:347}
  \b\mu = \frac1\mb\p\mu\phi + \delta_{\mu 0}\tilde{\b0}\quad\text{i.e.}\quad
  \left\{\begin{aligned}
    \bz &= \frac{\pz \phi}\mb\\
    \b0 & = \frac{\p0 \phi}\mb + \tilde{\b0}
  \end{aligned}\right.
\end{equation}
Notice that there is a residual uncertainty in the definition of $\phi$: one
can simultaneously shift $\phi\to \phi+ \beta(t)$ and $\tilde{\b0}\to
\tilde{\b0} -\frac{\dot \beta(t)}{\mb}$. Clearly, $\fboz = -\pz \tilde{\b0}$.

To establish the connection between $\phi$ and $\tilde{\b0}$ we consider the
Maxwell's equation:
\begin{equation}
  \label{eq:348}
  i \omega \fboz +\mb \pz \phi = -2\k \bar A_y(z) (i\omega \ax)\;,
\end{equation}
which gives
\begin{equation}
  \label{eq:349}
  \tilde{\b0} = \frac{\mb}{i\omega} \phi+2\k \int^z \bar A_y(z') \ax(z') dz'\;.
\end{equation}
Taking the derivative of eq.~(\ref{eq:248}) we arrive to the first class
constraint which massive vector fields \b\mu should obey
  \begin{equation}
  \label{eq:325}
  \p\mu B^\mu
  =-\frac{\k}{2\mb^2}\epsilon^{\mu\nu\lambda\rho}F_{\mu\nu}F_{\lambda\rho} =-
  \frac{4\k\h0 (i\omega \ax)}{\mb^2} 
\end{equation}
Putting relations~(\ref{eq:347}),~(\ref{eq:349}) into the
constraint~(\ref{eq:325}) we obtain
\begin{equation}
  \label{eq:350}
  (\Box +\mb^2)\phi = -\frac{4\k\h0 i\omega \ax}{\mb} - 2i\k\mb\omega \int^z
  \bar A_y(z') \ax(z') dz' 
\end{equation}
%
Next, we analyze the e.o.m. for the field \ax:
\begin{align}
  \label{eq:353}
  \Box \ax &= \frac{4\k\h0 i\omega}{\mb} \phi +
  \frac{4\k\h0\mb}{i\omega}\phi+8\k^2\h0 \int^z \bar A_y(z') \ax(z') dz' \\
  & \hphantom{=}+\frac{2\k\mb}{i\omega}\pz\phi\bar A_y + 4\k^2 \bar
  A_y^2\ax\notag
\end{align}
The propagation of light, polarized perpendicularly to the magnetic field
(component $A_y$), is not modified in this theory, as follows trivially from
the structure of currents in the right hand side of
eqs.~(\ref{eq:220})--(\ref{eq:248}).

Let us compare eqs.~(\ref{eq:350})--(\ref{eq:353}) to the pure axion case:
\begin{equation}
  \label{eq:244}
  \begin{aligned}
    \p\nu F^{\mu\nu} & = \frac1M \epsilon^{\mu\nu\lambda\rho} (\p\nu\phi)
    \f\lambda\rho\\
    (\Box +\ma^2)\phi &= \frac1{4M}\epsilon^{\mu\nu\lambda\rho}\f\mu\nu
    \f\lambda\rho
\end{aligned}
\end{equation}
For the propagation of linearly polarized light, parallel to the magnetic
field \h0 equations~(\ref{eq:244}) reduce to
\begin{equation}
  \label{eq:10}
  \begin{aligned}
    (\Box +\ma^2)\phi &= -\frac{2i\omega}{M} A_x \h0\\
    \Box A_x & = \frac{2i\omega\h0}{M}\phi
  \end{aligned}
\end{equation}
We see that under the identification 
\begin{equation}
  \label{eq:11}
  \frac1M\leftrightarrow \frac{2\k}\mb\quad;\quad \ma\leftrightarrow \mb
\end{equation}
equations~(\ref{eq:350})--(\ref{eq:353}) reduce to~(\ref{eq:10}) plus some
additional (non-local) terms, depending on the background potential $\bar
A_y$.\footnote{The term $ \frac{4\k\h0\mb}{i\omega}\phi$ in eq.~(\ref{eq:353})
  is subleading, as compared to the first term and does not change the results
  of our analysis in any essential way.}
These terms, however, are suppressed (as compared to those, depending on the
field \h0).  The easiest way to see it, is to send $\k\to0$, while keeping the
ratio $\k/\mb = \frac1{2M}$ \emph{fixed}. Thus,
equations~(\ref{eq:350})--(\ref{eq:353}) are reduced into those of massless
axion. This leads to the dichroism $\sim \frac{\k^2}{\mb^2}\h0^2 L^2$, where
$L$ is the distance traveled by the light. For finite \mb, corrections
suppressed by powers of $\frac\mb\omega$ will appear.


\section{Possible microscopic origin of the theory.}
\label{sec:avoid-stell-constr}

\subsection{Decoupling of the longitudinal polarization at high energies}
\label{sec:decoupling}

In quantum electrodynamics, if one adds a small photon mass \mg, all the
processes, involving the third (longitudinal) degree of freedom, are
suppressed at energies $E\gg \mg$ as $\frac\mg E$. This is due to the fact
that the electromagnetic field couples to a conserved current, which is a
consequence of gauge invariance of the theory.
On the other hand, if the current is not conserved, at high energies processes
involving the longitudinal polarization of the vector boson are equivalent to
those involving the longitudinally coupled scalar due to the so called
\emph{Goldstone boson equivalence theorem}~\cite{Cornwall:74}

This is what happens in theory~(\ref{eq:4}). Although the theory is gauge
invariant under the $U(1)_B$ gauge symmetry, it is realized by simultaneous
gauge transformations of the $B$-field and the Stuckelberg field $\theta$. As
we saw in Section~\ref{sec:prop-light-magn}, the field \b\mu couples to the
non-conserved current~(\ref{eq:7}) and therefore its longitudinal polarization
behaves as an axion (for $\omega \gg \mb$).

However, the theory~(\ref{eq:4}) is an effective field theory, valid up to a
certain energy scale $\mu$. In many cases it naturally happens that for $E\gtrsim
\mu$ the theory gets modified in such a way that the current, to which the
$\b\mu$-field couples becomes conserved. Then, all processes involving
emission or absorption of the longitudinal polarization of \b\mu are
suppressed as $\left(\frac\mb \omega\right)^2$.  On the other hand, we are interested in
the situation, where the field \b\mu can be produced at low laboratory energies.
This puts restriction on its mass to be $\mb \lesssim \omega_\lab\sim
1\ev$.  Then at stellar energies $\omega_\star \sim 1\kev$ one obtains a
suppression of \emph{at least} $\sim (\omega_\lab/\omega_\star)^2\sim
10^{-6}$. For smaller values of \mb\ the suppression is even
stronger.  The theory of transverse \b\mu-field with CS interaction does not
resemble the theory of axion anymore.  For example, the production of this
field due to the CS interaction is strongly suppressed by the small value of the
dimensionless CS coupling~$\k$.

To illustrate this idea, assume that in the theory there is an additional
particle with mass $\mf$, interacting with the fields of the
theory~(\ref{eq:4}) and giving rise to an effective action of the following
(schematic) form:
\begin{eqnarray}
  \label{eq:352}
  S = \int d^4x \left( -\frac14 F_A^2 -\frac14 F_B^2 +  \frac{\mb^2}2
    (D\theta^2)\right. \!\!\!\! &+&\!\!\!\! \k A\wedge B\wedge F_A \\  
\!\!\!\! &+&\!\!\!\! \k \left. \theta \frac{\mf^2}{\Box +
    \mf^2}(F\tilde F) - \k \,(\p\mu B^\mu) \frac{1}{\Box +
    \mf^2}(F\tilde F) \right)\nonumber
\end{eqnarray}
In the next subsection we will present an example of a renormalizable field
theory, which has such properties (recall that we always add to this theory
the constraint $F_A\wedge F_B = 0$ to make it gauge invariant). 
At \emph{low energies} (for $\omega < \mf$) one obtains the
action~(\ref{eq:4}) (formally taking $\mf\to\infty$).

At high energies ($\omega \gg \mf$) the effective theory is of course
non-local. We can neglect the interacting term proportional to $\theta$ in the
action~(\ref{eq:352}) and obtain
\begin{equation}
  \label{eq:6}
  S(\omega \gg \mf) \approx  \int -\frac14 F_A^2 -\frac14 F_B^2 +  \frac{\mb^2}2 D_\mu\theta^2 + \k A\wedge
  B\wedge F_A  - \k \,(\p\mu B^\mu) \frac{1}{\Box}(F\tilde F)
\end{equation}
At these energies, the field \b\mu couples to the \emph{conserved} current $J_B$
\begin{equation}
  \label{eq:8}
  J_B^\mu = \frac\k2 \epsilon^{\mu\nu\lambda\rho} A_\nu F_{\lambda\rho} -
  \k\frac{\p\mu}{\Box}(F\tilde F) \;.
\end{equation}
Therefore, at energies $\omega \gg \mf$ the production of the longitudinally
polarized \b\mu-field in theory~(\ref{eq:352}) is suppressed. Of course,
for $\omega>\mf$ the current~(\ref{eq:8}) should be computed directly in the
microscopic theory producing the non-local terms in~(\ref{eq:352}), containing 
additional particles, rather than
in the non-local effective theory. Next we are going to present an example of
such a microscopic theory.

\subsection{Theory with millicharged fermions}
\label{sec:full-model}

 \begin{table}[t]
  \centering
  \begin{tabular}[c]{
                     |>{$}c<{$}||>{$}c<{$}|>{$}c<{$}|>{$}c<{$}|>{$}c<{$}
                    ||>{$}c<{$}|>{$}c<{$}|>{$}c<{$}|>{$}c<{$}|>{$}c<{$}|} 
    \hline
    & \multicolumn{2}{c|}{$\Psi_1$} & \multicolumn{2}{c||}{$\Psi_2$} &
    \multicolumn{2}{c|}{$\Xi_1$} & \multicolumn{2}{c|}{$\Xi_2$} \\
    \hline
    &\psi_L & \psi_{R}& \psi_L' & \psi_{R}' &\chi_L &
    \chi_R & \chi_L' & \chi_R'\\
    \hline
    \u1_A 
& e_1 & e_1 & e_2 & e_2 & -e_3 +\delta{e} & e_3 & e_3 &
    -e_3 + \delta e \\
    \hline
    \u1_B & 
    q_{1} & -q_{1}  &  -q_1 & q_1 & q_2 & q_2 & q_2 & q_2
    \\\hline
  \end{tabular}
 \caption{A simple choice of charges of fermions, which leads
   to the low-energy effective action~(\ref{eq:352}). The charges are
chosen in such a way that all gauge anomalies cancel. The cancellation of
$U(1)_A^3$ and $U(1)_B^3$ anomalies happens for any value of
$e_i,q_i$. Cancellation of mixed anomalies requires $\delta e = \frac{\k}{4q_3
  e_3}$ where $\k$ is related to $e_i,q_i$ via~(\ref{eq:364}).} 
  \label{tab:charges}
\end{table}

Consider a theory with several sets of chiral fermions, whose masses are given
by Yukawa interactions with Higgs fields $\Phi_1$ and $\Phi_2$:
\begin{equation}
  \label{eq:12}
  \CL = \sum_{i=1,2}\Bigl(i\bar \Psi_i \Dsl \Psi_i + f_i \bar\Psi_i \Phi_1
  \Psi_i\Bigr) + \Bigl(i\bar \Xi_i \Dsl \Xi_i  + \lambda_i \bar\Xi_i \Phi_2
  \Xi_i\Bigr)+\mathrm{h.c.}
\end{equation}
The fermions are charged with respect to $U(1)_A\times U(1)_B$, with one of
the possible choices of charges shown in
Table~\ref{tab:charges}. 
%
%
Integrating out the fermions for energies below their masses, one obtains
``anomalous'' (CS-like) terms in the effective action
:\footnote{Of course, integrating out these fermions, one obtains also terms
  leading to a renormalization of charges of the fields $A_\mu$ and \b\mu, of
  the \b\mu mass, possible generation of kinetic mixing between $A_\mu$ and
  \b\mu, etc.  However, in the case of chiral fermions, one expects additional
  contributions from ``anomalous'' (triangular) diagrams (shown in
  Figs.~\ref{fig:diagram-triangular-theta}--\ref{fig:diagram-triangular}) (see
  e.g.~\cite{Anastasopoulos:06}).}
\begin{equation}
  \label{eq:1008}
  S_\cs =\int \frac{(e_1^2 -e_2^2)q_1}{16\pi^2} \theta F_A\wedge F_A + \k
  A\wedge  B\wedge F_A 
\end{equation}
Only fermions $\Psi$ contribute to the term $\theta F_A\wedge F_A$ (fermions
$\Xi$ are vector-like with respect to the $U(1)_B$ gauge group and therefore
do not interact with $\theta$). The contribution to the CS term $A\wedge
B\wedge F_A$ comes from both sets of fermions.  The interaction of the
\b\mu-field with the photon is very weak due to the small value of $\k$,
therefore in the action~(\ref{eq:1008}) we omitted anomalous terms, containing
more than one power of the $B$-field.

The relation between the coefficient $\k$ in front of the CS term and
the fermion charges is dictated by the gauge invariance of the action with respect to
$U(1)_B$ gauge transformation
\begin{equation}
\label{eq:364}
  \k = \frac{q_1 (e_1^2 - e_2^2)}{16\pi^2}
\end{equation}
\noindent 
The Higgs field $\Phi$ which provides mass to the fermions $\Psi_i$ is charged
with respect to the $U(1)_B$ with charge $2q_1$. If it acquires a vacuum
expectation value (VEV) $|\Phi_1| =\vb$, one has $\Phi_1 = \vb e^{2iq\theta}$
and the Yukawa term for $\Psi$ fields becomes $\vb\bar\Psi_i
e^{2iq_1\theta\g5} \Psi_i$. The Higgs then provides mass to the \b\mu-field:
\begin{equation}
  \label{eq:375}
  \mb = 2 q_1 \vb
\end{equation}
and Higgs's kinetic term becomes that of the field $\theta$ in
action~(\ref{eq:4}).

\begin{figure}[t]
  \centering
  \includegraphics{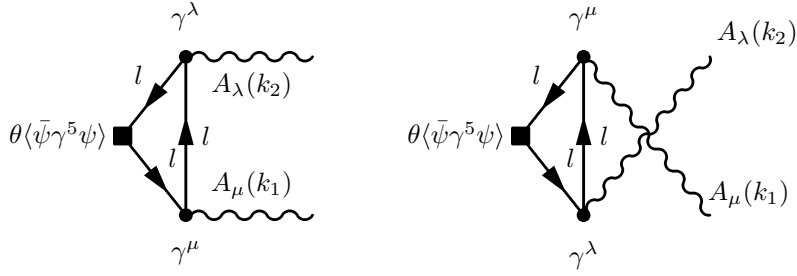} %
  \caption{Anomalous contributions to the correlator $\la\bar \psi \gamma^5 \psi\ra$
.} %
  \label{fig:diagram-triangular-theta}
\end{figure}

\begin{figure}[t]
  \centering %
  \includegraphics{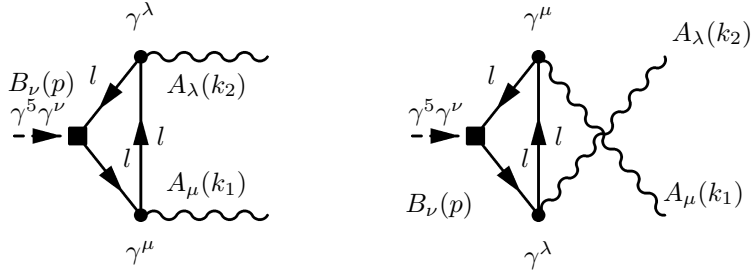}
  \caption{Two graphs, contributing to the Chern-Simons terms} %
  \label{fig:diagram-triangular} %
\end{figure}

Let us take Yukawa couplings $f_i \ll \lambda_i$, so that fermions $\Xi$ are
much heavier that $\Psi$. Consider energies $m_\Psi < E < m_\Xi$.  
The resulting expression for the current $J_B^\mu$ is given by the expression
\begin{equation}
  \label{eq:13}
  J^\mu_B = \frac\k2 \epsilon^{\mu\nu\lambda\rho} A_\nu F_{\lambda\rho} -
  \k\frac{\p\mu}{\Box}(F\tilde F)
\end{equation}
which arises from the effective action
\begin{equation}
  \label{eq:22}
  S_{eff,high} = \int \k(\p\mu B^\mu)\frac 1{\Box} F_A\wedge F_A + \k
    A\wedge B\wedge F_A
\end{equation}
(compare e.g.~\cite{Smailagic:00} where similar result for the massless chiral
fermions is presented).

\subsection{The parameters of the model}
\label{sec:numbers-1}

Next, we study the restrictions on the parameters of our model. First of all,
notice that in order to expect positive outcome in laboratory experiments,
which aim to measure dichroism and birefringence or photon regeneration
(shining light through the wall), we need to have a mass of the \b\mu-field to
be within the energy reach of laboratory experiments:
\begin{equation}
  \label{eq:15}
  \mb \lesssim \omega_{\lab} \sim 1\ev
\end{equation}
This means that
\begin{equation}
  \label{eq:16}
  \vb \lesssim 1\ev/q_1
\end{equation}
Notice, that the charges of fermions with respect to the ``paraphoton''
$B_\mu$ are not restricted and therefore one may have $q_1 \lesssim 1$. This
will make \vb\ to be also in the sub-eV region.

When speaking about ALPs, it is convenient to characterize their interaction
with the photon by the dimensionful coupling $M$ (see eqs.~(\ref{eq:244})).
It follows from eq.~(\ref{eq:11}) that in terms of the parameters of our model
$M \sim \frac\mb\k$. Using then relations~(\ref{eq:364}) and (\ref{eq:375}),
we obtain the following expression for the interaction strength $M$ in terms
of the parameters of the microscopic model:\footnote{We consider the case when
  $e_1 \sim e_2\sim \qf$.}
\begin{equation}
  \label{eq:14}
  M \sim \frac{\vb}{\qf^2}
\end{equation}
For our analysis of the classical equations of motion to remain valid, the
tree-level unitarity bound for processes involving massive vector bosons
should not be saturated at all relevant energies. The saturation of this
bound is reached at energies $E\sim \mb/\k$ which is again $M$.

There are various restrictions on the charges of new types of fermions with
the electromagnetic field. First, laboratory bounds, coming from the
contributions to the Lamb shift~\cite{Davidson:00} and invisible
orthopositronium decay~\cite{Davidson:00} (based on the results
of~\cite{Mitsui:93}) give $\qf < 10^{-4}$. This translates into
\begin{equation}
  \label{eq:18}
  M \gtrsim 10^8 \vb = 0.1\gev\parfrac{\vb}{\!\ev}
\end{equation}
This bound is very weak.  A stronger bound on the charges of millicharged
fermions ($\qf <10^{-6}$) with sub-eV masses comes from the requirement that
such fermions do not distort the CMB spectrum too much~\cite{Melchiorri:07}.
However, this restriction is not applicable in our case as the mass of our
fermions $\mf > 1\ev$~(see discussion below).

The strongest bounds on charges of fermions with mass below $\sim 30\kev$
comes from limiting the contribution of these particles to the energy transfer
in stars~\cite{Davidson:91} (see also~\cite{Raffelt-book}). This gives $\qf
<10^{-14}$. This bound translates into the following lower bound on $M$:
\begin{equation}
  \label{eq:19}
  M > 10^{28}\vb = 10^{19}\gev\parfrac{\vb}{\!\ev}
\end{equation}
For instance, if we take $M\sim 10^9\gev$ (the sensitivity expected to be
reached by the OSQAR experiment~\cite{OSQAR:06}), this would require $\vb\sim
10^{-10}$. This would imply an extremely light \b\mu field with $\mb \sim
10^{-10}\ev$ and $\k \sim 10^{-28}$ (which is still a possibility).

However, the mass \mb\ is not necessarily so small. In our model, the
paraphoton field \b\mu would acquire a kinetic mixing with the photon due to
the loop corrections coming from light fermions. Therefore, the mechanism of
additional suppression of the coupling of fermions with the photon in stars,
similar to that proposed in~\cite{Masso:06a}, is possible. The restriction
then becomes $\qf \lesssim 10^{-14} \parfrac{\omega_\star}{\mb}^2$, i.e. the
stellar bound of~\cite{Davidson:91,Davidson:00} is weakened by \emph{at least}
six orders of magnitude. This would lead to the following bound
\begin{equation}
  \label{eq:20}
  M \gtrsim 10^7\gev\parfrac{\vb}{\!\ev}^5
\end{equation}
with $\mb \lesssim 1\ev$.

We are interested in the situation where at laboratory energies the effective
action is given by~(\ref{eq:9}). For this to be true, the mass $m_\Psi$ of the
fermions $\Psi$ should be greater than laboratory energies $\omega_\lab$. At
the same time, the mass should be lower than the stellar energies
$\omega_\star \sim 1\kev$ for~(\ref{eq:13}) to be true. Therefore the fermion
mass \mf\ lies in the interval $ 1\ev \lesssim m_\Psi \lesssim 1\kev$. As we
saw in Section~\ref{sec:decoupling}, in the limit $\mf/\omega_\star\to0$
the massive vector field \b\mu is coupled to a conserved current and therefore
its production in stars is suppressed \emph{at least} as
$\bigl({\omega_\lab}/{\omega_\star}\bigr)^2$.  At higher energies ($\omega
\sim \omega_\star$) additional contributions to the action~(\ref{eq:22}) due
to finite $m_\Psi/\omega_\star$ behave as
$\bigl({m_\Psi}/\omega_\star\bigr)^2$. These terms contribute to the
divergence of the current $J_B^\mu$:
\begin{equation}
  \label{eq:29}
  \p\mu J^\mu_B \sim \k \parfrac{\mf}{\omega_\star}^2 \epsilon^{\mu\nu\lambda\rho} F_{\mu\nu}F_{\lambda\rho}
\end{equation}
This gives an additional contribution to the production of the longitudinally
polarized $B$-field, not suppressed by $(\mb/\omega)^2$. However, in
comparison with the low-energy expression~(\ref{eq:7}), there is an additional
suppression $\bigl({m_\Psi}/\omega_\star\bigr)^2$. For convenience, in what
follows, we will characterize the interaction strength in terms of an energy
dependent coupling constant $M(\omega)$:
\begin{equation}
  \label{eq:30}
  M(\omega) \sim \frac{\mb}{\k}\parfrac{\omega}{\mf}^2\quad,\quad\omega> \mf
\end{equation}

To determine the admissible values of \mf, let us assume that one of the
laboratory experiments (light propagation in a magnetic field or
regeneration experiments) had positive outcome. This determines the value of
$M_\lab\sim \mb/\k$. As a conservative estimate we take it to be equal to the
laboratory lower bound, reported by PVLAS collaboration~\cite{PVLAS:07a}:
$M_\lab \gtrsim 2\times 10^6\gev$. This value is much lower than those
coming from the astrophysical bounds: restriction from the helioseismology
$M_\odot \gtrsim 2\times 10^9\gev$~\cite{Schlattl:98,Bahcall:04} and
restrictions from the horizontal branch (HB) stars: $M_\text{HB} \gtrsim
10^{10}\gev$ (for a review see e.g.~\cite{Raffelt-book,Raffelt:99,Raffelt:06a}).
Demanding that the modification~(\ref{eq:30}) at star energies
$M(\omega_\star)$ is greater than these astrophysical bounds (which we will
collectively call $M_\star$), one arrives to the following restriction on the
mass \mf:
\begin{equation}
  \label{eq:23}
  m_\Psi \lesssim \omega_\star\sqrt{\frac{M_\lab}{M_\star}}
\end{equation}
As an estimate for $\omega_\star$ we take the plasma frequency in the stellar
interim: $\omega_\star \sim 0.3 \kev$ for the Sun and $\omega_\star \sim
2\kev$ for HB stars. As a result, we obtain $\mf \lesssim 10\ev$ using solar data
or $\mf \lesssim 20\ev$ using restrictions from HB stars.

The recent bound from CAST collaboration~\cite{CAST:07a,CAST:07b}
$M_\text{CAST} \gtrsim 10^{10}\gev$ would give the restriction on the mass of
the fermions $\mf \lesssim$ few eV. However, it should be noted that the
present CAST bounds do not extend above axion masses $\sim 0.02 \ev$. The
model presented here allows $\mb \gtrsim 0.02\ev$ (while still below
$\omega_\lab$), therefore the CAST restrictions may not be applicable in our
case.\footnote{The CAST Phase II will test the range of masses $0.02 \ev < m_a
  < 1.1\ev$~\cite{CAST:07b} and will allow to probe further our model.}
  
Additional restriction on the parameter space may come from the following
fact.  Integrating out fermions $\Psi$ leads in general to terms suppressed by
$1/\mf$ (as in any renormalizable field theory~\cite{Appelquist:74}). At
laboratory experiment energies with $\omega \sim 1\ev$, such contributions may
still be significant.  In particular, they contribute to the action the
following terms (analogous to the Euler-Heisenberg corrections in quantum
electrodynamics~\cite{Heisenberg:36})
\begin{equation}
  \label{eq:371}
  S_\Psi = \int d^4 x\, \frac2{45}\frac{\qf^4}{\mf^4}\left(
    \frac{7}{16}\ffd^2 + 4 (F_{\mu\nu}^2)^2\right)
\end{equation}
These terms of course do not contribute to dichroism (as we consider $\omega <
2\mf$), while their contribution to birefringence is given by
\begin{equation}
\label{eq:366}
  \beta_\psi \sim (\omega L) \frac{\qf^4}{\mf^4}\h0^2
\end{equation}
Taking $\qf \sim 10^{-8}$, $\mf \sim 1\ev$ (to maximize the value
of~(\ref{eq:366})), and typical values for $\omega \sim 1\ev$, $L\sim 1$~m, and
$\h0 \sim 5$~T, one finds a value of birefringence $\beta_\Psi \sim 10^{-20}$ --
much below the current experimental limits.

Thus, the final range of allowed parameters of millicharge fermions ranges
from
\begin{equation}
  \label{eq:17}
  \vb \lesssim 1\ev\quad; \quad \qf \lesssim 10^{-8}\quad ;\quad q_\Psi < 1
\end{equation}
to
\begin{equation}
  \label{eq:21}
   \vb \lesssim 10^{-10}\ev\quad; \quad \qf \lesssim 10^{-14}\quad ;\quad
   q_\Psi < 1
\end{equation}
with the fermion mass $\omega_\lab\lesssim \mf < 20\ev$.\footnote{The allowed
  window for \mf\ can become higher, if future laboratory experiments
  strengthen bounds of~\cite{PVLAS:07a}.}

We see, that this range of parameters differs from those of
Refs.~\cite{Gies:06a,Ahlers:06}, which suggested that the PVLAS experiment can
be explained by the existence of millicharged particles with $\mf \sim 0.1\ev$
and charges $\qf \sim 10^{-6}$. Such millicharged particles, however, would
introduce a distortion of CMB energy spectrum~\cite{Melchiorri:07}.  On the
other hand, the millicharged particles with
parameters~(\ref{eq:17})--(\ref{eq:21}) and vector field with
parameters~(\ref{eq:20}) can give rise to effects testable in present and
forthcoming experiments (PVLAS, OSQAR, ALPS, LIPPS, BMV, Q\&A): shining light
through the wall and dichroism and birefringence of light propagating in
strong magnetic field.\footnote{Note that as eq.~(\ref{eq:20}) shows, taking
  \vb\ in the sub-eV range allows to have $M\sim 10^5\gev$ and thus explaining
  the effect of dichroism, reported by the PVLAS collaboration in
  2006~\cite{PVLAS:05a}.  To avoid the stellar constraints this would require
  to have $\mf\sim \omega_\lab\sim 1\ev$. However, the contribution of these
  fermions to dichroism at laboratory energies is subleading compared to the
  corresponding contribution of the CS term~(\ref{eq:9}), which differs this
  model from the one proposed in~\cite{Ahlers:06,Gies:06a}.}

\section{Discussion}
\label{sec:discussion}

In this paper we presented a simple example of a theory in which non-trivial
anomaly cancellation between light and heavy sectors gives rise to possible
observable effects in optical laboratory experiments. We showed that the
structure of the effective action changes at high energies, suppressing the
production of light particles. Therefore the stellar constraints can be
significantly weaker in this case, allowing for relatively strong optical
effects, that may be observed in current or future experiments.

We would like to notice that although we have presented an explicit microscopic
theory~(\ref{eq:12}), which leads to the effective theory~(\ref{eq:352}) we are interested
in, the actual fundamental theory could be rather different. For example, instead of heavy
set of fermions $\Xi$ in the theory~(\ref{eq:12}), the CS term may be generated by different
mechanisms in the underlying high-energy theory. 

Actually, the low-energy effective action~(\ref{eq:4}) or~(\ref{eq:9}) with
the CS coupling can arise easily in D-brane realizations of the Standard
Model~\cite{Antoniadis:02,Anastasopoulos:06}.  Indeed, gauge groups come often
in unitary factors containing extra $U(1)$'s.  Moreover, the smallness of the
mass $m_B$ and the CS coupling $\k$ of the vector boson $B_\mu$ to the photon
may be naturally explained in models of large extra dimensions and low string
scale~\cite{ArkaniHamed:98,Antoniadis:98}.  For this, $B$ should propagate in
the bulk of extra dimensions, while its mass should arise from localized
anomalies due to chiral fermions present for instance in the intersections of
the Standard Model branes with the brane extended in the bulk.  Assuming for
simplicity an homogeneous bulk of volume $v_\perp$ in string units and
identifying the axion with the string scales $M\sim M_s$, one has $\k\sim
1/\sqrt{v_\perp}$ and $m_B\sim M/\sqrt{v_\perp}$. It follows that $m_B\sim 1$
eV for $M\sim 100$ TeV, in which case $\k\simeq 10^{-14}$. On the other hand,
the microscopic theory with millicharged fermions is more difficult to
accommodate.  One could imagine for instance that the hypercharge (or the
electric charge) acquires a tiny mixing with the bulk gauge boson $B$,
creating millicharges, although it is not clear how to get a consistent setup
with ``natural" suppression.

Another example of a theory with non-trivial anomaly cancellation between
light and heavy sectors, is in the presence of extra dimensions,
discussed in~\cite{anomaly-th,anomaly-exp}. In this theory a 
plane wave, propagating in a strong magnetic field $H_x \approx
  \mathbf{const}$ with polarization parallel to the field, is described
  by the equation
\begin{equation}
  \frac1{\Delta(z)}\p z \Bigl(\Delta(z) \p z A_x\Bigr) + \Box A_x =
  \frac{\alpha_\text{EM}^2 \kappa_0^2 \vec H^2}{M_5^2\Delta^2(z)}
  A_x+ \mathcal{O} (\kappa_0)
\end{equation}
This leads to the massive wave equation
\begin{equation}
\label{eq:25}  
  \Box A_x(x) - \mh^2\,A_x(x) = 0
\end{equation}    
where the ``magnetic mass''
\begin{equation}
  \label{eq:26}
  \mh^2 \sim \alpha_{\textsc{em}}\kappa_0 |\vec H|  
\end{equation}
depends only on 4-dim quantities; it is not suppressed by the scale of the 5-dim
theory $M_5$. The wave equation for perpendicular to the magnetic field
component remains the same (massless): $ \Box A_y(x) =0$.

This theory is drastically different from the models discussed in this paper
(as well as other models, predicting non-trivial effects in strong magnetic
fields). Namely, it does not contain any new light degrees of freedom. This
means, that the regeneration experiments (``shining light through the wall''),
as well as the measurements of dichroism will produce no results in this case.
However the theory leads to an {ellipticity} (birefringence) of the linearly
polarized light:
\begin{equation}
  \label{eq:28}
  \Delta \phi = \frac{\mh^2}{2\omega}L\sim
  \frac{\kappa_0 \alpha_\text{EM} |\vec H|}{2\omega} L
\end{equation}
Actually, the static (``capacitor'') experiment suggested in~\cite{anomaly-exp} is also
sensitive to the model described in this paper. As different models,
predicting non-trivial results in optical experiments behave differently in
this static experiment, it provides a good way to distinguish among them. 

\section*{Acknowledgements}

We would like to thank M.~Fairbairn, G.~Raffelt and M.~Shaposhnikov for useful
discussions.  This work was supported in part by the European Commission under
the RTN contract MRTN-CT-2004-503369 and in part by the INTAS contract
03-51-6346.  The work of A.B. was (partially) supported by the EU 6th
Framework Marie Curie Research and Training network "UniverseNet"
(MRTN-CT-2006-035863). O.R. would like to thank Swiss Science Foundation.

\providecommand{\href}[2]{#2}\begingroup\raggedright%

\endgroup

\end{document}